\documentclass[aps,pra,twocolumn,showpacs]{revtex4-1}

\usepackage{amsfonts,amssymb,amsmath, braket}
\usepackage{mathtools}
\usepackage[]{graphics,graphicx,epsfig}
\usepackage{amsthm,multirow}
\usepackage{color}
\usepackage{diagbox}
\begin{document}

\title{Bound for Gaussian-state Quantum illumination using direct photon measurement}

\author{Su-Yong \surname{Lee}}
\email[]{suyong2@add.re.kr}

\author{Dong Hwan Kim}
\email[]{kiow639@gmail.com}

\author{Yonggi Jo}

\author{Taek Jeong}



\author{Duk Y. Kim}

\author{Zaeill Kim}
\affiliation{Advanced Defense Science \& Technology Research Institute,Agency for Defense Development, Daejeon 34186, Republic of Korea}

\date{\today}

\begin{abstract}
It is important to find feasible measurement bounds for quantum information protocols.
We present analytic bounds for quantum illumination with Gaussian states when using an on-off detection or a photon number resolving (PNR) detection, where its performance is evaluated with signal-to-noise ratio. First, for coincidence counting measurement, the best performance is given by the two-mode squeezed vacuum (TMSV) state which outperforms the coherent state and the classically correlated thermal (CCT) state. However, the coherent state can beat the TMSV state with increasing signal mean photon number in the case of the on-off detection. Second, the performance is enhanced by taking Fisher information approach of all counting probabilities including non-detection events.
In the Fisher information approach, the TMSV state still presents the best performance but the CCT state can beat the TMSV state with increasing signal mean photon number in the case of the on-off detection. Furthermore, we show that it is useful to take the PNR detection on the signal mode and the on-off detection on the idler mode, which reaches similar performance of using PNR detections on both modes.
\end{abstract}

\maketitle

\section{Introduction}
Entanglement is indispensable for quantum teleportation, quantum sensing, and quantum illumination (QI). 
Contrary to the other protocols, QI loses entanglement but takes quantum advantage by survival of quantum correlation.
The objective of QI is to discriminate the presence or absence of a low-reflectivity target \cite{Lloyd,Tan}. 
According to the frequency range of the probe signal used for QI, photon loss is the dominant limiting factor for the performance in optical wave range or thermal noise is the dominant one in microwave range. In a laboratory, a low-reflectivity beam splitter plays the role of the target and thermal noise is intentionally injected into the low-reflectivity beam splitter.
Thermal noise is replaced by thermal state which is produced by scattering coherent state light with a rotating glass disk or blocking one mode of a two-mode squeezed vacuum (TMSV) state.
We consider a scenario where the signal mode of an input state interacts with a target having a low reflectivity in a strong thermal-noise environment, and then the reflected signal mode is measured in a receiver. Given an idler mode of the input state, it is best to measure the idler mode with the reflected signal mode.

The performance of QI can be evaluated by quantum Chernoff bound (QCB) \cite{Aud07,Calsa08,Stefano08}, which is the upper limit of the lower bound on a target-detection error probability.
Under a weak thermal-noise environment, in the beginning, an entangled state takes quantum advantage over a separable state under single-photon level \cite{Lloyd}. 
The idea was extended to Gaussian states under a strong thermal-noise environment  \cite{Tan,Athena}, 
where a TMSV state outperforms a coherent state. 
As an input state, it is feasible to prepare Gaussian states, such as coherent, thermal, squeezed, and TMSV states.
Based on QCB, coherent state presents the best performance in single-mode Gaussian states.
In two-mode Gaussian states, TMSV state is a nearly optimal state \cite{Ranjith,Bradshaw}  for symmetric discrimination and an optimal state \cite{Palma} for asymmetric one.
A classically correlated thermal (CCT) state, which is produced by impinging a thermal state into a beam splitter, cannot outperform the TMSV state and  the coherent state in QCB.

Although QCB is not directly related to any physical observable, it can be achieved with signal-to-noise ratio (SNR) with specific measurement schemes. 
QCB and SNR are independently derived from the detection error probability. 
The QCB is represented by the exponential of the decay constant \cite{Aud07,Kim}, $\exp[-M\gamma]$, as the upper limit of the lower bound, and
the SNR is represented by the exponent of the detection error probability, $\exp[-\text{SNR}^{(M)}]$, where $M$ is the number of modes. 
As examples,
coherent state asymptotically approaches its QCB with homodyne detection \cite{Tan,SL09}, 
and CCT state can do that with photon number difference measurement \cite{Lee22}.
However, TMSV state asymptotically can attain its QCB  not by the SNR that only considers joint local measurement but by collective measurement, e.g., sum frequency generation with feedforward \cite{Quntao}, which requires a quantum memory.
Except the homodyne detection, the two other measurement schemes were not implemented in a laboratory.

Since the first QI experiment \cite{Genovese}, there were several experiments implemented in optical range \cite{Zheshen,England,Aguilar,Sussman,Xu} and microwave range \cite{Sandbo,Luong,Shabir19, Assouly}.
Theoretically, there are feasible proposals with optical parametric amplifier (OPA) and phase-conjugate (PC) receivers \cite{Guha}, photon number difference measurement \cite{Lee,Noh, Lee22}, homodyne (or heterodyne) detection \cite{Blakely,Jo,Gaetana}, and on-off detection scheme \cite{Jeffers,Jeffers22} that is conditionally to prepare a signal mode by detecting the idler mode with an on-off detector (or on-off detector arrays).
Note that the OPA and PC receivers can asymptotically approach a half of the exponent of the QCB for TMSV state.
However, all the measurements schemes require nonlinear interactions or interference with additional modes, except the on-off detection and the photon number difference measurement.
Thus, it is worthwhile to find the detection error probability bound using direct photon measurement, without any other additional modes.


In this paper, we consider the most feasible measurements for Gaussian-state quantum-illumination, using on-off detection and photon number resolving (PNR) detection. 
Note that we have already considered homodyne (or heterodyne) detection in our previous work \cite{Lee22}.
Physically, the on-off detection can be implemented by an ideal avalanche photodiode (APD) as well as an ideal superconducting nanowire single-photon detector (SNSPD), where both detectors are commercially available. The PNR detection can be implemented by an ideal transition-edge-sensor (TES) that is extensively investigated \cite{Pfister} as well as SNSPD arrays \cite{Divochiy}.
CCD or EMCCD is also implemented to discriminate photon numbers, even with low efficiency. 
Theoretically, it is valuable to find the best theoretical bounds using the on-off and PNR detection, without additional modes or combining the signal-and-idler mode.
By directly measuring the idler and the reflected signal, we evaluate the performance bound with signal-to-noise ratio (SNR).
The performance is shown by coincidence counting events, and then it is enhanced by Fisher information approach including all possible counting events. 

\section{Single- and two-mode Gaussian states}
A Gaussian state is described with covariance matrix and first-order moments \cite{Weedbrook}. 
The covariance matrix is described with $\sigma_{jk}=\langle \hat{R}_j \hat{R}_k +\hat{R}_k \hat{R}_j\rangle -2\langle \hat{R}_j\rangle \langle \hat{R}_k\rangle$, where $\hat{R}_k=\hat{X}_k (\hat{P}_k)$ and $\hat{a}_k=\frac{1}{\sqrt{2}}(\hat{X}_k+i\hat{P}_k)$. 
The first-order moments are described with displacement, $\mu^T=\sqrt{2}(\text{Re}(\alpha),~  \text{Im}(\alpha))$.
$\hat{a}_k$ is the annihilation operator in mode $k$ that can be described with the position and momentum operators, $\hat{X}_k$  and $\hat{P}_k$. According to the relation among the annihilation operator, the position and momentum operators, the coefficients in the covariance matrix and the first-order moment are determined, such as $2$ in $\sigma_{jk}$ and $\sqrt{2}$ in $\mu^T$.
After interacting a signal mode with a target in thermal noise environment, 
the covariance matrix is transformed into $\sigma(\kappa)=X\sigma_{\text{in}}X^T+Y$, where $X=\text{diag}(\sqrt{\kappa},\sqrt{\kappa},1,1), 
Y=\text{diag}(1-\kappa+2N_B, 1-\kappa+2N_B, 0, 0),$ and $\sigma_{\text{in}}$ is the input covariance matrix. 
The first-order moment is transformed into $\mu(\kappa)=X\mu_{\text{in}}$, where $\mu_{\text{in}}$ is the input first-order moment.
$\kappa$ is a target reflectivity while $1-\kappa$ is the target transmittivity.

Previously, most of the QI works compared the performances of TMSV state and coherent state, while there are few works that compared the performances of TMSV state and CCT state \cite{Genovese,Sandbo,Luong}. Recently one compared the performances of coherent state, CCT state, and TMSV state \cite{Shabir19,Lee22}.
Here we consider coherent state, CCT state, TMSV state, and displaced squeezed (DS) state.

A single-mode pure Gaussian state is represented by a DS state $\hat{D}(\alpha)|\xi\rangle$, 
where $\alpha=|\alpha|e^{i\phi}$ and $\xi=r e^{i\varphi}$.
 $\alpha$ is a displacement parameter, $\xi$ is a squeezing parameter, and $r$ is the amplitude of the squeezing parameter.
After interacting the input signal with a target having reflectivity $\kappa$ in thermal noise environment, the reflected output state is given by
\begin{align}
	 \sigma_{\text{DS}}(\kappa)=\begin{pmatrix}
 A_1-A_2\cos{\varphi}  & -A_2\sin{\varphi} \\
-A_2\sin{\varphi}  & A_1+A_2\cos{\varphi}
	\end{pmatrix}, \label{DS}
\end{align}
\begin{align}
~\mu=\sqrt{2\kappa}|\alpha|\begin{pmatrix}
\cos{\phi}\\
\sin{\phi}
\end{pmatrix},\nonumber
\end{align}
where $A_1=1+2N_B+2\kappa N_{sq}$, $A_2=2\kappa\sqrt{N_{sq}(N_{sq}+1)}$, $N_{sq}=\sinh^2 r$, and $N_B$ is the mean photon number of thermal noise observed at a detector. 
When the target is absent, the covariance matrix is $\sigma_{\text{DS}}(0)$ and the first-order moment is $\mu=0$.

We also consider two-mode squeezed vacuum (TMSV) states and classically correlated thermal (CCT) states as representatives of two-mode Gaussian states with no first-order moment.
The former is a representative of continuous variable entangled states, and the latter is a kind of classically correlated states.
After the signal mode of the TMSV state interacts with a target, the output is given by
\begin{align}
	 \sigma_{\text{TMSV}}(\kappa)=\begin{pmatrix}
 B & 0 & C & 0 \\
0 & B & 0 & -C \\
C & 0 & 1+2N_S & 0\\
0 & -C &0 & 1+2N_S
	\end{pmatrix},
\end{align}
where $B=1+2N_B+2\kappa N_S$, and $C=2\sqrt{\kappa N_S(N_S+1)}$.
 $N_S$ is the mean photon number of the signal mode.
When the target is absent,  the covariance matrix is $\sigma_{\text{TMSV}}(0)$.
After the signal mode of the CCT state interacts with the target, the output state is given by
\begin{align}
	 \sigma_{\text{CCT}}(\kappa)=\begin{pmatrix}
 B & 0 & D & 0 \\
0 & B & 0 & D \\
D & 0 & 1+2N_I & 0\\
0 & D &0 & 1+2N_I
	\end{pmatrix},
\end{align}
where $D=2\sqrt{\kappa N_S N_I}$ and $N_I$ is the mean photon number of the idler mode.
When the target is absent, the covariance matrix is $\sigma_{\text{CCT}}(0)$.
The pre-interaction covariance matrices of the TMSV and CCT states are obtained by writing down Eqs. (2) and (3) 
at $\kappa= 1$ and $N_B = 0$.

\section{Signal-to-Noise Ratio with coincidence counting}
The performance of discriminating the presence or absence of a target can be evaluated with
a detection error probability that is represented by a sum of miss-detection probability $P(\text{off}|\text{on})$ and false-alarm probability $P(\text{on}|\text{off})$, 
which is minimized under the decision threshold. 
Assuming unbiasedness for two hypotheses, such as the presence and absence of a target, we take equal
prior probabilities for both miss-detection and false-alarm probabilities. Then, the detection error probability is given by 
$P^{(M)}_{\text{err}}=\frac{1}{2}[P(\text{off}|\text{on})+P(\text{on}|\text{off})]$ whose minimum is approximately upper bounded as 
$P^{(M)}_{\text{err}}\approx \exp[-\text{SNR}^{(M)}]$ when the two states of target present or absent are close to each other \cite{Guha,Lee22,Chernoff} at $M\gg1$.
Under mode-by-mode measurements, the signal-to-noise ratio (SNR) is explicitly described with
\begin{align}
\text{SNR}^{(M)}=\frac{M(\langle \hat{O}\rangle_{\kappa} - \langle \hat{O}\rangle_{\kappa=0})^2}{2[\sqrt{\Delta^2 O_{\kappa}}+\sqrt{\Delta^2 O_{\kappa=0}}]^2}, \label{SNRM}
\end{align}
where $M$ is the number of modes, $\langle \hat{O}\rangle_{\kappa}$ is the mean value of an observable $\hat{O}$, and $\Delta^2 O_{\kappa}$ is its variance.
Increasing the SNR corresponds to decreasing the detection error probability.

\subsection{On-off detection}
An observable that describes coincident on-off detection is given by $\hat{O}_{\text{on}}\equiv\bigotimes_{i=S,I}(\hat{I}_i-|0\rangle_{i}\langle 0|)$ for signal and idler modes, where $\hat{I}_i$ is the identity operator of mode $i$. 
With no idler mode, it only takes the signal-mode observable. Since the variance of the observable is given as
$\Delta^2 O_{\text{on}}=\langle \hat{O}_{\text{on}}\rangle (1-\langle \hat{O}_{\text{on}}\rangle)$, 
the SNR of Eq.~(\ref{SNRM}) is represented only with $\langle \hat{O}_{\text{on}}\rangle$.

For a single-mode Gaussian state, the mean value of the observable is obtained as 
$\langle \hat{O}_{\text{on}}\rangle=1-tr(\rho_{\text{out}}|0\rangle_S\langle 0|)$. Given  a DS state, the vacuum probability from the Eq.~(\ref{DS}) is derived as
\begin{align}
\langle 0|\rho_{\text{out}}|0\rangle=\frac{\exp[-\frac{\kappa |\alpha|^2}{2L}(A_1+1+A_2\cos(\varphi-2\phi))]}{\sqrt{L}}, 
\label{off_single}
\end{align}
where $L=(N_B+1)^2+\kappa N_{sq}(2+2N_B-\kappa)$. Then, the SNR is maximized at $\cos(\varphi-2\phi)=1$.
At $\kappa=0.01$ and $N_B=600$, the SNR is maximized with $|\alpha|^2\approx 0.918$ in the constraint of $N_S\equiv|\alpha|^2+N_{sq}=1$,
where the DS state presents sub-Poissonian statistics.
Since there is a very small difference between the SNRs of the coherent state and the DS state, we choose the coherent state as the representative of the single-mode Gaussian state in the on-off detection scheme.
In the limit of $N_S,~\kappa \ll 1\ll N_B$, the SNR is approximated as $\frac{M\kappa^2 N_S^2}{8N_B^3}$.

For a two-mode Gaussian state, the mean value of the observable is obtained as 
$\langle \hat{O}_{\text{on}}\rangle=1-tr(\rho_{\text{out}}|0\rangle_S\langle 0|)-tr(\rho_{\text{out}}|0\rangle_I\langle 0|)+
tr(\rho_{\text{out}}|00\rangle_{SI}\langle 00|)$. Given a TMSV state, the mean value of the observable is derived as
\begin{eqnarray}
\label{off_TMSV}
\langle \hat{O}_{\text{on}}\rangle=\frac{N_S(1+N_B)+1}{(1+N_S)(1+N_B)}-\frac{1}{1+N_B+\kappa N_S}. 
\end{eqnarray}
In the limit of $N_S,~\kappa \ll 1\ll N_B$, the SNR is approximated as $\frac{M\kappa^2 N_S}{8N_B^4}$.
Given a CCT state, the mean value of the observable is derived as
\begin{eqnarray}
\label{off_CCT}
\langle \hat{O}_{\text{on}}\rangle&=&1-\frac{1}{1+N_B+\kappa N_S}-\frac{1}{1+N_I}\\
&&+\frac{1}{(1+N_I)(1+N_B)+\kappa N_S}.\nonumber
\end{eqnarray}
In the limit of $N_S,N_I,~\kappa \ll 1\ll N_B$, the SNR is approximated as $\frac{M\kappa^2 N_S^2 N_I}{2N_B^4}$.

\begin{figure}[ht!]
\includegraphics[width=0.45\textwidth]{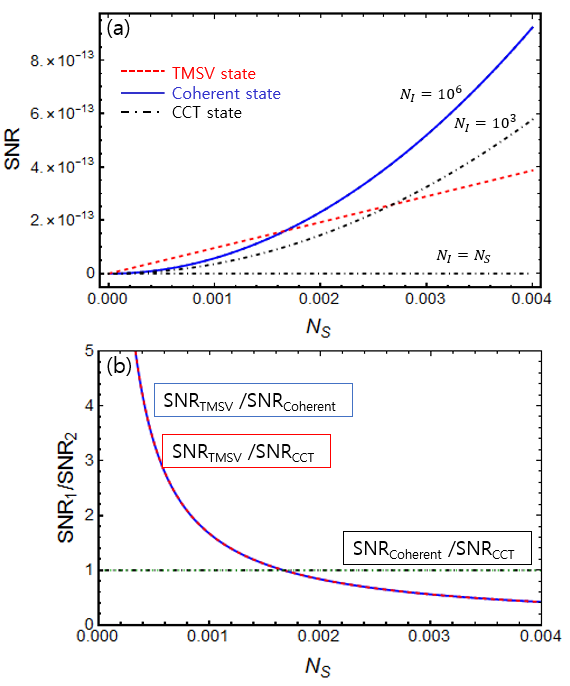}
\caption{(a) Signal-to-noise ratio as a function of $N_S$ under coincidence counting measurements using on-off detection: TMSV state (red, dashed), coherent state (blue, solid), and CCT state (black, dot-dashed).
According to the mean photon number of the idler mode ($N_I$) in the CCT state, the CCT state can beat the TMSV state and approach the performance of the coherent state at $N_I=10^6$.
(b) SNR ratios for TMSV/Coherent (blue, solid), TMSV/CCT  (red, dashed), and Coherent/CCT (black, dot-dashed) states at $N_I=10^6$. 
Blue and red curves are overlapped. Green dotted line indicates $1$.  In all cases $\kappa=0.01$, $N_B=600$, and $M=10^6$.
}
\label{fig:fig1}
\end{figure}

Fig.\ref{fig:fig1}(a) shows the SNRs of the three Gaussian states as a function of $N_S$ under coincidence counting measurements using on-off detectors. 
The TMSV state outperforms the coherent state and the CCT state at $N_S< 0.0016$, but
the coherent state presents the best performance at $N_S >0.0016$. 
The TMSV state cannot take advantage of quantum correlation over coherent state in a high mean photon number of the signal mode when using on-off detection.
The CCT state can beat the TMSV state with increasing the mean photon number of the idler mode $N_I$, and then it can approach the coherent state performance around $N_I=10^6$.
Increasing the mean photon number of the idler mode of the CCT state corresponds to increasing the classical correlation of the CCT state. Under the on-off detections, increasing the classical correlation of the CCT state can beat the performance of the TMSV state for a fixed signal mean photon number.
In Fig.\ref{fig:fig1}(b), we show the SNR ratios between the states.
The TMSV state can take advantage over the coherent state more than five times at low $N_S$.
The SNR ratio between the coherent  state and the CCT state with $N_I=10^6$ is equal to one, regardless of $N_S$.

\subsection{Photon Number Resolving detection}
Replacing the on-off detector by the PNR detector, it is expected to extract more information encoded in different photon number states.
An observable of coincident PNR detection is given by $\hat{n}_{SI}\equiv\bigotimes_{i=S,I}\hat{a}^{\dag}_i\hat{a}_i$.
It only takes the signal mode observable for a single-mode state.

For a single-mode Gaussian state, the mean value of the observable is obtained as $\langle \hat{n}_S\rangle=\kappa\langle \hat{n}_S\rangle_{\text{in}}+N_B$, where subscript `$\text{in}$' represents the input mode. 
 $\hat{n}_S$ represents the number operator of the signal mode.
Its variance is given by 
$\Delta^2 n_s=\kappa \langle \hat{n}_S\rangle_{\text{in}} (1+2N_B+\kappa Q_M)+N_B(1+N_B)$, where $Q_M\equiv\frac{\langle \hat{n}^2_S\rangle_{\text{in}}-\langle \hat{n}_S\rangle^2_{\text{in}}}{\langle \hat{n}_S\rangle_{\text{in}}}-1$ is the Mandel Q-factor.
The SNR of Eq.~(\ref{SNRM}) increases with decreasing $Q_M$. 
Thus, it is best to send a single-mode Gaussian state having sub-Poissonian statistics, i.e., DS state, toward the target.
In the range of $\kappa\ll 1$, however, it is negligible to consider the sub-Poissonian contribution from the $\kappa^2$ factor, such that it is much more feasible to prepare the coherent state than the DS state.
In the limit of  $N_S,~\kappa \ll 1\ll N_B$, the SNR is approximated as $\frac{M\kappa^2 N_S^2}{8N_B^2}$.

For a two-mode Gaussian state, the mean value of the observable is obtained as 
$\langle \hat{n}_{SI}\rangle=\kappa \langle \hat{n}_S\hat{n}_I\rangle_{\text{in}}+N_B\langle \hat{n}_I\rangle_{\text{in}}$.
 $\hat{n}_I$ represents the number operator of the idler mode.
Its variance is given by
\begin{eqnarray}
\Delta^2 n_{SI}&=&\kappa^2(\langle (\hat{n}_S \hat{n}_I)^2 \rangle_{\text{in}}-\langle \hat{n}_S \hat{n}_I \rangle^2_{\text{in}})
+N_B \langle \hat{n}^2_I\rangle_{\text{in}}\\
&&+N_B^2(2\langle \hat{n}^2_I\rangle_{\text{in}}-\langle \hat{n}_I\rangle^2_{\text{in}})-2\kappa N_B \langle \hat{n}_S\hat{n}_I\rangle_{\text{in}} \langle \hat{n}_I\rangle_{\text{in}} \nonumber\\
&&+\kappa(1-\kappa +4N_B)\langle \hat{n}_S\hat{n}^2_I \rangle_{\text{in}}. 
\nonumber
\end{eqnarray}
In the limit of $N_S,~N_I,~\kappa \ll 1\ll N_B$, 
the SNR of Eq.~(\ref{SNRM}) is approximated as
\begin{eqnarray}
\text{SNR}^{(M)}\approx \frac{M\kappa^2\langle \hat{n}_S \hat{n}_I \rangle^2_{\text{in}}}{8[N_B^2(2\langle \hat{n}^2_I\rangle_{\text{in}}-\langle \hat{n}_I\rangle^2_{\text{in}})+N_B\langle \hat{n}^2_I\rangle_{\text{in}}]}.
\end{eqnarray}
Given a TMSV state and a CCT state, the SNRs are approximated as $\frac{M\kappa^2 N_S}{16N_B^2}$ 
and $\frac{M\kappa^2 N_S^2N_I}{4N_B^2}$, respectively.

\begin{figure}
\includegraphics[width=0.45\textwidth]{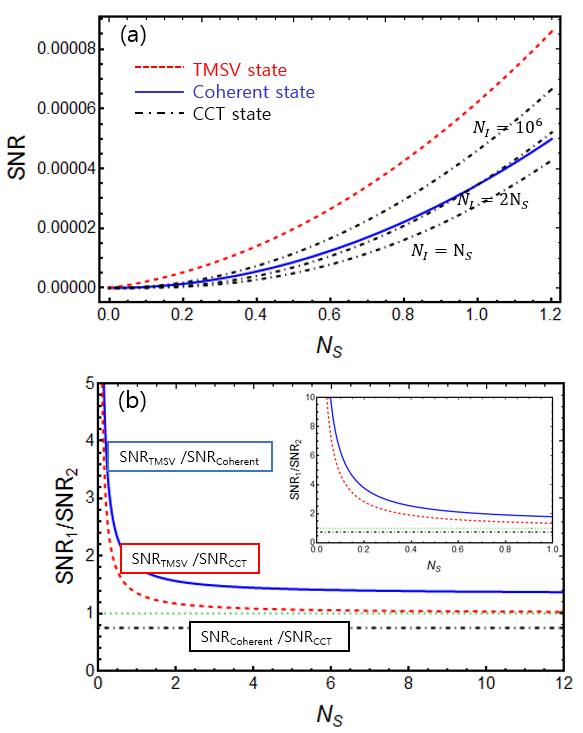}
\caption{(a) Signal-to-noise ratio as a function of $N_S$ under coincidence counting measurements using PNR detections: TMSV state (red, dashed), coherent state (blue, solid), and CCT state (black, dot-dashed).
According to the mean photon number of the idler mode ($N_I$) in the CCT state, the CCT state can beat the coherent state.
(b) SNR ratios for TMSV/Coherent (blue, solid), TMSV/CCT (red, dashed), and Coherent/CCT (black, dot-dashed) states at $N_I=10^6$.
Green dotted line indicates $1$. In all cases $\kappa=0.01$, $N_B=600$, and $M=10^6$.
}
\label{fig:fig2}
\end{figure}

Fig.\ref{fig:fig2}(a) shows the SNRs of the three Gaussian states as a function of $N_S$ under coincidence counting measurements using PNR detectors.  The TMSV state outperforms the other states in the whole range of $N_S$. 
The TMSV state takes advantage of quantum correlation over coherent state even in a high mean photon number of the signal mode when using the PNR detection. It is understood that the PNR detection can extract much more information on the target as well as on quantum correlation of the TMSV state than the on-off detection.
With increasing $N_I$, the CCT state can beat the coherent state at $N_I=2N_S$ but it cannot beat the TMSV state.
Under PNR detections, increasing the classical correlation of the CCT state cannot beat the performance of the TMSV state for a fixed signal mean photon number.
In Fig.\ref{fig:fig2}(b), we show the SNR ratios between the states.
The TMSV state can take advantage over the coherent (or CCT) state more than ten times at low $N_S$.
The SNR ratio between the coherent  state and the CCT state with $N_I=10^6$ is less than one, regardless of $N_S$.

\section{Signal-to-Noise Ratio with Fisher information approach}
In the limit of $\kappa\ll 1$, the SNR of Eq.~(\ref{SNRM}) can be approximated  as quantum Fisher information (QFI)  \cite{Noh, Sanz} that is achieved with symmetric logarithmic derivative (SLD) \cite{Paris}. The eigenbasis of the SLD corresponds to the optimal measurement basis.
Putting an optimal quantum estimator $\hat{O}_{\eta}=\eta\hat{I}+\hat{L}_{\eta}/H_{\eta}$ of the Ref. \cite{Paris} into the SNR of Eq. (4), the SNR is approximated as 
$\text{SNR}^{(M)}\approx M{\kappa} H_{(\kappa = 0)}/8$, where $\kappa=\eta^2$, $H_{\kappa}$ is the QFI, and $\hat{I}$ is the identity operator.
The SLD is given by $\hat{L}_{\kappa}=2\sum_{n,m}\frac{\langle \psi_m|\partial_{\kappa} \rho_{\kappa}|\psi_n\rangle}{p_n+p_m}|\psi_m\rangle\langle\psi_n |$, where $\rho_{\kappa}=\sum_n p_n|\psi_n\rangle\langle \psi_n |$.

Under a specific measurement scheme described by an orthogonal POVM, the outcome statistics can be encoded in a diagonal density matrix whose diagonals are the probabilities of each outcome.
The SLD of the diagonal density matrix describes the optimal observable that can be performed under the specific measurement.
As the measurement basis is diagonalized, the associated quantum estimator contains only diagonal terms so that the associated SLD is given by
\begin{align}
\hat{L}_{D}=\sum_n x_n\hat{M}_n \equiv \hat{O}_{pt},~
x_n = \left. \frac{\partial_\kappa \text{Tr}(\rho_\kappa \hat{M}_n)}{\text{Tr} (\rho_\kappa \hat{M}_n)} \right|_{\kappa=0},
\end{align} 
where $\partial_{\kappa}$ represents the partial derivative with respect to $\kappa$.
 Here, $\{\hat{M}_n\}$ represents the orthogonal POVM corresponding to either on-off or PNR detection scheme.
Putting the associated SLD into the SNR of Eq. (4), we can obtain $\langle\hat{L}_D\rangle_{\kappa} \approx \kappa F_{(\kappa= 0)}$ and 
 $\langle\hat{L}^2_D\rangle_{\kappa} -  \langle\hat{L}_D\rangle^2_{\kappa} \approx F_{(\kappa= 0)}$.
Thus, we can derive the SNR being approximated as 
\begin{align}
\text{SNR}^{(M)}\approx \frac{M\kappa^2}{8}\sum_n\frac{[\partial_{\kappa}P_n]^2}{P_n}\bigg|_{\kappa=0}
=\frac{M\kappa^2}{8}F_{(\kappa= 0)},\label{FI}
\end{align}
 where $P_n=\text{Tr} (\rho_\kappa \hat{M}_n)$ is the probability and $F_{(\kappa= 0)}$ is the Fisher information.  
Note that we take a coherent state as the representative of the single-mode Gaussian state, due to its negligible difference from the DS state.


\subsection{On-off detection}
FI using on-off detection is represented by $\sum_{i,j=\text{on,~off}}\frac{[\partial_{\kappa} P_{oo}(i,j|\kappa)]^2}{P_{oo}(i,j|\kappa)}|_{\kappa=0}$.
$P_{oo}(i,j|\kappa)$, which is a function of the target reflectivity $\kappa$, represents the conditional probability of presenting $i$ status in the reflected mode and $j$ status in the idler mode.	
For a coherent state, the SNR is approximated as $\frac{M\kappa^2 N_S^2}{8N_B(N_B+1)^2}$ by using Eq.~(\ref{off_single}).
For a TMSV state, the SNR is approximated as $\frac{M\kappa^2 N_S(1+N_S)}{8N_B(N_B+1)^2}$ by using Eq.~(\ref{off_TMSV}).
For a CCT state, the SNR is approximated as $\frac{M\kappa^2 N_S^2}{8N_B(N_B+1)^2}[1+\frac{N_I}{(1+N_I)^2}]$ by using Eq.~(\ref{off_CCT}), which is maximized at $N_I=1$. 
The TMSV state outperforms the other states in the range of $N_S < 4$, but the CCT state presents the best performance at $N_S>4$.
By using the on-off detection, such as $\hat{\Pi}_0=|0\rangle\langle 0|$ and $\hat{\Pi}_1=\hat{I}-\hat{\Pi}_0$,
the corresponding SLDs are obtained as follows: $\hat{O}_{pt}=N_B\hat{\Pi}_{0,S}-\hat{\Pi}_{1,S}$ for the coherent state,
$\hat{O}_{pt}=(N_B\hat{\Pi}_{0,S}-\hat{\Pi}_{1,S})\otimes \hat{\Pi}_{1,I}$ for the TMSV state, and
$\hat{O}_{pt}=(N_B\hat{\Pi}_{0,S}-\hat{\Pi}_{1,S})\otimes (\hat{\Pi}_{0,I}+(2+N_I)\hat{\Pi}_{1,I})$ for the CCT state.
The subscript S (or I) represents signal (or idler) mode.

\subsection{Photon Number Resolving detection}
FI using PNR detection is represented by $\sum_{n,m}\frac{[\partial_{\kappa} P(n,m|\kappa)]^2}{P(n,m|\kappa)}|_{\kappa=0}$,
where $P(n,m|\kappa)$ is the conditional probability of detecting $n$ photons in the reflected mode and $m$ photons in the idler mode.
The probability $P(n,m|\kappa)$ can be calculated 
by using the Bargmann representation in Appendix A.
For the coherent state, the SNR is approximated as $\frac{M\kappa^2 N^2_S}{8N_B(N_B+1)}$.
For the TMSV state, the SNR is approximated as $\frac{M\kappa^2 N_S(1+2N_S)}{8N_B(N_B+1)}$.
For the CCT state, the SNR is approximated as $\frac{M\kappa^2 N^2_S(1+2N_I)}{8N_B(N_B+1)(1+N_I)}$, which is maximized at $N_I\rightarrow \infty$.
The TMSV state presents the best performance in the whole range of $N_S$. 
Note that the coherent state cannot beat the other states in both on-off detection and PNR detection.
By using the PNR detection,
the corresponding SLDs are obtained as follows: 
$\hat{O}_{pt}=N_B\hat{I}_S-\hat{n}_S$ for the coherent state, 
$\hat{O}_{pt}=(N_B\hat{I}_S-\hat{n}_S)\otimes \hat{n}_I$ for the TMSV state, and
$\hat{O}_{pt}=(N_B\hat{I}_S-\hat{n}_S)\otimes (\hat{n}_I+\hat{I}_I)$ for the CCT state.
The subscript S (or I) represents signal (or idler) mode.

\begin{table}[h!]
\centering
\begin{tabular}{|c || c | c| c |}
\hline
SNR/ Input state&
DS state&
CCT state&
TMSV state \\[0.2ex]
\hline
 & & & \\
CC (on-off detection)&  $\frac{M\kappa^2 N^2_S}{8N^3_B}$ & $\frac{M\kappa^2 N^2_SN_I}{2N^4_B}$ &  
$\frac{M\kappa^2 N_S}{8N^4_B}$ \\
 & & & \\
FI (on-off detection)&  $\frac{M\kappa^2 N^2_S}{8N^3_B}$ & $\frac{M\kappa^2 N^2_S}{8N^3_B}$ &  
$\frac{M\kappa^2 N_S}{8N^3_B}$ \\
 & & & \\
CC (PNR detection)&  $\frac{M\kappa^2 N^2_S}{8N^2_B}$ & $\frac{M\kappa^2 N^2_SN_I}{4N^2_B}$ &  
$\frac{M\kappa^2 N_S}{16N^2_B}$ \\
 & & & \\
FI  (PNR detection)&  $\frac{M\kappa^2 N^2_S}{8N^2_B}$ & $\frac{M\kappa^2 N^2_S}{8N^2_B}$ &  
$\frac{M\kappa^2 N_S}{8N^2_B}$ \\
 & & & \\
\hline
\end{tabular}
\caption{SNR at $N_S,N_I,\kappa \ll 1 \ll N_B$.  
CC represents coincidence counting. The coherent state is measured with a PNR detection or an on-off detection on a signal mode. }
\label{table:1}
\end{table}

In Table \ref{table:1}, we compare the SNRs that are derived from coincidence counting as well as FI in the limit of $N_S,N_I,\kappa \ll 1 \ll N_B$.
Since on-off detection and PNR detection measure the diagonal components of the output states, the scaling of the SNRs is proportional to $\kappa^2$.
For each state, the PNR detection presents better performance than the on-off detection, 
and FI approach outperforms the coincidence counting approach.
Since non-coincidence counting events also include target information, it is utilized to enhance the SNR by the FI approach.
Coherent state presents the same ordering of SNRs between the FI and coincidence counting approaches.
The SNR of the CCT is enhanced by the FI approach as much as $N_B/4N_I$ times for the on-off detection and $1/2N_I$ times for the PNR detection.
It also enhances the SNR of the TMSV as much as $N_B$ times for the on-off detection and twice for the PNR detection.

\section{Combining PNR and on-off detections}

\begin{figure}
\includegraphics[width=0.45\textwidth]{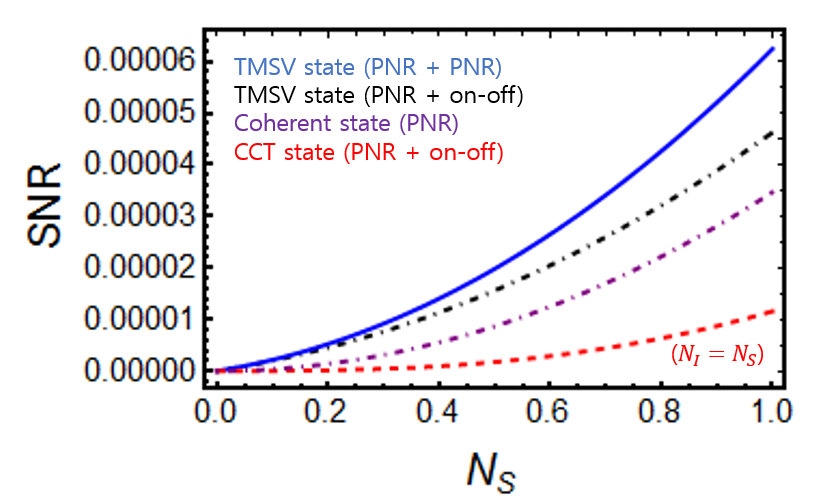}
\caption{ Signal-to-noise ratio as a function of $N_S$ under coincidence counting measurements, where $\kappa=0.01$, $N_B=600$, and $M=10^6$.
 TMSV state using PNR detection on both modes (blue, solid), TMSV state using PNR detection on the signal mode and on-off detection (black, dot-dashed),
 Coherent state using PNR detection on the signal mode (purple, dot-dashed), and  CCT state at $N_S=N_I$ using PNR detection on the signal mode and on-off detection (red, dashed).
With increasing the mean photon number of the idler mode ($N_I$) in the CCT state, the CCT state can approach the performance of the coherent state using the PNR detection.
}
\label{fig:fig3}
\end{figure}

As a further step, we can combine PNR detection and on-off detection in a single measurement setup, such as PNR detection on the signal mode and on-off detection on the idler mode, or vice versa. The former observable is given by $\hat{M}_{no}=\hat{n}_S\otimes (\hat{I}_I-|0\rangle_I\langle 0|)$ 
and the latter one is given by $\hat{M}_{on}=(\hat{I}_S-|0\rangle_S\langle 0|)\otimes \hat{n}_I $. 
Since the performance does not take advantage when using on-off detection on the signal mode and PNR detection on the idler mode, we present only the case of the PNR detection on the signal mode and on-off detection on the idler mode.
Target information is encoded in the signal mode such that it is more informative to discriminate the photon numbers of the signal mode than of the idler mode. Even if the idler mode is measured with PNR detection, the performance is significantly affected by the signal mode measurement setup. The case of (PNR+on-off) considers the number of photons on signal mode when clicking on the idler mode for each signal-idler pair. For Fisher information approach, we consider even the non-click events, such as (0+0), (m+0), and (0+m) (m is a positive integer).
Fig.3 shows the SNRs of the three Gaussian states as a function of $N_S$ under coincidence counting measurements.
We show that, for TMSV state,  the case of the PNR detection on a signal mode and the on-off detection on an idler mode can reach similar performance of using PNR detection on both modes. 
It can be explained by the ability of extracting a target information. After a signal mode interacts with a target, the target information is encoded in a reflected signal mode. Then, it is more informative to extract the target information with high photon number discrimination capability by measuring the reflected signal mode than measuring the idler mode. On the other hand, it does not show a dramatic improvement of the performance by discriminating more than two photons in the idler mode. Thus, although on-off detection is considered in the idler mode, the (PNR+on-off) detection can exhibit the similar performance of the (PNR+PNR) detection.

Given the TMSV state, the moments of the observable are derived as 
\begin{eqnarray}
\langle \hat{M}_{no}\rangle&=&\kappa N_S+\frac{N_SN_B}{1+N_S},~\\
\langle \hat{M}^2_{no}\rangle&=&(\kappa N_S+N_B)(1+2\kappa N_S+2N_B)-\frac{N_B(2N_B+1)}{1+N_S}.\nonumber
\end{eqnarray}
In the limit of $N_S,\kappa \ll 1\ll N_B$, the SNR scales as $M\kappa^2N_S/(16N^2_B)$,
which implies that it is enough to prepare a PNR detection on the signal mode and an on-off detection on the idler mode in order to reach similar performance of using PNR detection on both modes.
We also obtain the similar behavior with the Fisher information approach, where both cases exhibit the same scaling of $M\kappa^2N_S/(8N^2_B)$.

\section{Summary and Discussion}
We have presented analytically the performance bound of Gaussian-state quantum illumination using on-off detection or PNR detection, 
without any additional operation on the idler and the reflected signal.
Under coincidence counting measurements, we showed that TMSV state presents the best performance, 
whereas coherent state can beat the TMSV state with increasing $N_S$ when using on-off detection.
Moreover, we presented that the performance of the TMSV state can be enhanced by taking FI approach with all counting probabilities including non-detection events, whereas CCT state can beat the TMSV state with increasing $N_S$ when using on-off detection. 
We also found that DS state having sub-Poissonian statistics exhibits the best performance in the single-mode Gaussian state even if the advantage is very small compared to the coherent state. 
Furthermore, we showed that the performance of taking PNR detection on a signal mode and the on-off detection on an idler mode exhibits similar performance of using PNR detection on both modes.



Direct photon detection is the simplest scheme of measuring the correlation between the signal and idler modes, which does not require additional modes to interfere with the signal (or idler) mode. 
Since it cannot measure off-diagonal components of output states that provide quantum correlation,
the direct photon detection cannot take quantum advantage over the classical limit given by coherent state.
Under the same detection scheme, however, we showed that TMSV state can take quantum advantage over the coherent (or CCT) state.
As a further work, we may consider a practical scenario, e.g., detection efficiencies, where we would obtain the crossover between SNRs of TMSV state and coherent (or CCT) state for different input parameter conditions.
For a simplest detection bound, we expect that the direct photon measurement scheme can be applied to secure quantum communication \cite{Stefano20,Candia21},  long (or short)-range object sensing, and quantum reading \cite{SP}, which is based on discriminating quantum states or quantum channels.

\begin{acknowledgments}
This work was supported by a grant to Defense-Specialized Project funded by Defense Acquisition Program Administration and Agency for Defense Development.

\end{acknowledgments}

\onecolumngrid
\section*{Appendix A: Bargmann representation}
The Bargmann representation of a state is a compact method to encode the number basis matrix elements in a single function \cite{Hall}. For example, the Bargmann representation of a one-mode state $\rho$ is a function $K(z,w^*)$ in two variables $z$, $w^*$, and the matrix element $\braket{n|\rho |m}$ is given as $ \sqrt{n!m!}\times(\text{coeff. of } z^n w^{*m})$. Hence by computing the Bargmann representation of the returned states, we have access to all number state probabilities. 

The reflected output state when using a displaced squeezed state has Bargmann representation
\begin{align}
\begin{split}
K_{\text{DS}}(z,w^*) &= \frac{1}{ \sqrt{L}} \text{exp} \left[ \frac{1}{2L} \left\{ \frac{1}{2}(A_1^2-A_2^2-1) z w^* -A_2 (e^{i\varphi} z^2 +e^{-i\varphi} w^{*2})+\sqrt{\kappa} (\alpha(A_1+1)+\alpha^* e^{i\varphi}A_2) z \right. \right.\\
&\hspace{25mm}\left. \left.+ \sqrt{\kappa}(\alpha^* (A_1+1)+\alpha e^{-i\varphi}A_2) w^* - \kappa |\alpha|^2 (A_1+1+A_2 \cos(2\phi-\varphi)) \right\} \right],\end{split}
\end{align}
where $A_1=1+2N_B+2\kappa N_{sq}$, $A_2=2\kappa\sqrt{N_{sq}(N_{sq}+1)}$, $N_{sq}=\sinh^2 r$, and $L=(N_B+1)^2+\kappa N_{sq}(2+2N_B-\kappa)$.

The reflected output state when using a TMSV state has Bargmann representation
\begin{align}
K_{\text{TMSV}}(z_1, z_2, w_1^*, w_2^*) &= \frac{1}{ E} \text{exp} \left[\frac{2(N_{S}+1)(B-1) - C^2}{4E} z_1 w_1^* + \frac{2N_{S}(B+1) - C^2}{4E} z_2 w_2^*   +\frac{C}{2E}(z_1 z_2 +  w_1^* w_2^*)  \right],
\end{align}
where $B=1+2\kappa N_S+2N_B$, $C=2\sqrt{\kappa N_S(N_S+1)}$, and $E=(N_B+1)(N_{S}+1)$.

The reflected output state when using a CCT state has Bargmann representation
\begin{align}
K_{\text{CCT}}(z_1, z_2, w_1^*, w_2^*) &= \frac{1}{ F} \text{exp} \left[ \frac{2(N_{I}+1)(B-1) - D^2}{4F} z_1 w_1^* + \frac{2N_{I}(B+1) -D^2}{4F} z_2 w_2^*   +\frac{D}{2F}(z_1 w_2^* + w_1^* z_2 ) \right],
\end{align}
where $D=2\sqrt{\kappa N_S N_I}$ and $F= (N_B+1)(N_{I}+1)+\kappa N_S$.




\begin{thebibliography}{99}

\bibitem{Lloyd} S. Lloyd,  Science {\bf 321}, pp. 1463--1465 (2008).

\bibitem{Tan} S.-H. Tan, B. I. Erkmen, V. Giovannetti, S. Guha, S. Lloyd, L. Maccone, S. Pirandola, and J. H. Shapiro,  Phys. Rev. Lett. {\bf 101}, 253601 (2008).



\bibitem{Aud07} K.M.R. Audenaert, J. Calsamiglia, R. Munoz-Tapia, E. Bagan, L. Masanes, A. Acin, and F. Verstraete, 
\prl \textbf{98},160501 (2007).

\bibitem{Calsa08} J. Calsamiglia, R. Munoz-Tapia, L. Masanes, A. Acin, and E. Bagan, 
\pra \textbf{77}, 032311 (2008).

\bibitem{Stefano08} S. Pirandola and S. Lloyd,
\pra \textbf{78}, 012331 (2008).

\bibitem{Athena} A. Karsa, G. Spedalieri, Q. Zhuang, and S. Pirandola, Phy.Rev.Research \textbf{2}, 023414 (2020).


\bibitem{Ranjith} R. Nair and M. Gu, 
Optica \textbf{7}, 771(2020).

\bibitem{Bradshaw} M. Bradshaw, L.O. Conlon, S. Tserkis, M. Gu, P.K. Lam, and S.M. Assad, 
\pra \textbf{103}, 062413 (2021).

\bibitem{Palma} G. De Palma and J. Borregaard, 
\pra \textbf{98}, 012101 (2018).


\bibitem{Kim} D.H. Kim, Y. Jo, D.Y. Kim, T. Jeong, J. Kim, N.H. Park, Z. Kim, and S.-Y. Lee, 
Phys.Rev.Research \textbf{5},033010 (2023).

\bibitem{SL09} J.H. Shapiro and S. Lloyd, 
New J. Phys. \textbf{11}, 063045 (2009).

\bibitem{Lee22} S.-Y. Lee, Y. Jo, T. Jeong, J. Kim, D.H. Kim, D. Kim, D.Y. Kim, Y.S. Ihn, and Z. Kim, 
\pra \textbf{105}, 042412 (2022).

\bibitem{Quntao} Q. Zhuang, Z. Zhang, and J.H. Shapiro, 
\prl \textbf{118}, 040801 (2017).

\bibitem{Genovese} E.D. Lopaeva, I. Ruo Berchera, I.P. Degiovanni, S. Olivares, G. Brida, and M. Genovese, 
\prl \textbf{110}, 153603 (2013).

\bibitem{Zheshen} Z. Zhang, S. Mouradian, F.N.C. Wong, and J.H. Shapiro, 
\prl \textbf{114}, 110506 (2015).

\bibitem{England}D.G. England, B. Balaji, and B.J. Sussman, 
\pra \textbf{99}, 023828 (2019).

\bibitem{Aguilar} G.H. Aguilar, M.A. de Souza, R.M. Gomes, J. Thompson, M. Gu, L.C. Celeri, and S.P. Walborn, 
\pra \textbf{99}, 053813 (2019). 

\bibitem{Sussman} Y. Zhang, D. England, A. Nomerotski, P. Svihra, S. Ferrante, P. Hockett, and B. Sussman, 
\pra \textbf{101}, 053808 (2020).

\bibitem{Xu} F. Xu, X.-M. Zhang, L. Xu, T. Jiang, M.-H. Yung, and J. Zhang, 
\prl \textbf{127}, 040504 (2021).

\bibitem{Sandbo}C.W. Sandbo Chang, A. M. Vadiraj, J. Bourassa, B. Balaji, and C.M. Wilson, 
Appl. Phys. Lett. \textbf{114}, 112601 (2019).

\bibitem{Luong} D. Luong, C. W. S. Chang, A. M. Vadiraj, A. Damini, C. M. Wilson, B. Balaji,  
IEEE Trans. Aero. Electron. Syst. \textbf{56},2041 (2020).

\bibitem{Shabir19} S. Barzanjeh,  S. Pirandola, D. Vitali, and J.M. Fink, 
Sci. Adv. \textbf{6}, eabb0451 (2020).

\bibitem{Assouly} R. Assouly, R. Dassonneville, T. Peronnin, A. Bienfait, and B. Huard, 
Nature physics (2023), http://doi.org/10.1038/s41567-023-02113-4.

\bibitem{Guha}S. Guha and B.I. Erkmen, 
\pra \textbf{80}, 052310 (2009).


\bibitem{Lee} S.-Y. Lee, Y.S. Ihn, and Z. Kim, 
\pra \textbf{103}, 012411 (2021).

\bibitem{Noh} C. Noh, C. Lee, and S.-Y. Lee, 
J. Opt. Soc. Am. B \textbf{39}, 1316 (2022).




\bibitem{Blakely} J.N. Blakely, 
Quantum Rep. \textbf{2}, 400 (2020).

\bibitem{Jo} Y. Jo, S. Lee, Y.S. Ihn, Z. Kim, and S.-Y. Lee, 
Phys.Rev.Research \textbf{3}, 013006 (2021).

\bibitem{Gaetana}G. Spedalieri and S. Pirandola, 
Phys.Rev.Research \textbf{3}, L042039 (2021).





\bibitem{Jeffers} H. Yang, W. Roga, J.D. Pritchard, and J. Jeffers, 
Opt. Express \textbf{29}, 8199 (2021).

\bibitem{Jeffers22} H. Yang, N. Samantaray, and J. Jeffers, 
Phys. Rev. Applied \textbf{18}, 034021 (2022).

\bibitem{Pfister} M. Eaton, A. Hossameldin, R.J. Birrittella, P.M. Alsing, C.C. Gerry, H. Dong, C. Cuevas, and O. Pfister, 
Nature Photon. (2022), 
https://doi.org/10.1038/s41566-022-01105-9.

\bibitem{Divochiy} A. Divochiy et al., 
Nature Photon. \textbf{2}, 302 (2008).

\bibitem{Weedbrook} C. Weedbrook, S. Pirandola, R. Garcia-Patron, N.J. Cerf, T.C. Ralph, J.H. Shapiro, and S. Lloyd, 
\rmp \textbf{84}, 621 (2012).




\bibitem{Chernoff} H. Chernoff, 
Ann. Math. Statist. \textbf{23}, 493 (1952).


\bibitem{Sanz} M. Sanz, U. Las Heras, J.J. Garc\'ia-Ripoll, E. Solano, and R. Di Candia, 
\prl \textbf{118}, 070803 (2017).

\bibitem{Paris} M.G.A. Paris, 
Int. J. Quantum Inf. \textbf{7}, 125 (2009).



\bibitem{Stefano20} S. Pirandola, U.L. Andersen. L. Banchi, M. Berta, D. Bunandar, and R. Colbeck et al., 
Adv. Opt.  Photon \textbf{12}, 1012 (2020).

\bibitem{Candia21} R. Di Candia, H. Yigitler, G.S. Paraoanu, and R. Jantti, 
PRX Quantum \textbf{2}, 020316 (2021).

\bibitem{SP} S. Pirandola, 
\prl \textbf{106}, 090504 (2011).


\bibitem{Hall} B. C. Hall,  
Contemporary Mathematics, \textbf{260}, 1-59 (2000).






\end{thebibliography}
\end{document}